\documentclass[sigconf]{acmart}

\setcopyright{acmcopyright}
\copyrightyear{2022}
\acmYear{2022}
\acmDOI{}

\acmConference[ICPC 2022]{The 30th International Conference on Program Comprehension}{May 21–22, 2022}{Pittsburgh, PA, USA}
\acmBooktitle{ICPC '22: IEEE/ACM International Conference on Program Comprehension,
  May 21--22, 2022, Pittsburgh, Pennsylvania, United States}
\acmPrice{15.00}
\acmISBN{978-1-4503-XXXX-X/18/06}

\usepackage{csquotes}
\usepackage{booktabs,tabularx,makecell}
\usepackage[inline]{enumitem}
\begin{document}

\title[Anchoring Code Understandability Evaluations]{Anchoring Code Understandability Evaluations Through Task Descriptions}

\author{Marvin Wyrich}
\email{marvin.wyrich@iste.uni-stuttgart.de}
\orcid{0000-0001-8506-3294}
\affiliation{%
  \institution{University of Stuttgart}
  \city{Stuttgart}
  \country{Germany}
  \postcode{70569}
}

\author{Lasse Merz}
\email{lasse.merz@web.de}
\affiliation{%
  \institution{University of Stuttgart}
  \city{Stuttgart}
  \country{Germany}
  \postcode{70569}
}

\author{Daniel Graziotin}
\email{daniel.graziotin@iste.uni-stuttgart.de}
\orcid{0000-0002-9107-7681}
\affiliation{%
  \institution{University of Stuttgart}
  \city{Stuttgart}
  \country{Germany}
  \postcode{70569}}

\renewcommand{\shortauthors}{Wyrich et al.}

\newcommand{\totalNumbParticipants}{256 }
\newcommand{\numbStudentParticipants}{206 }
\newcommand{\numbProfessionalParticipants}{50 }

\begin{abstract}
  In code comprehension experiments, participants are usually told at the beginning what kind of code comprehension task to expect. Describing experiment scenarios and experimental tasks will influence participants in ways that are sometimes hard to predict and control. In particular, describing or even mentioning the difficulty of a code comprehension task might anchor participants and their perception of the task itself. \\ In this study, we investigated in a randomized, controlled experiment with \totalNumbParticipants participants (\numbProfessionalParticipants software professionals and \numbStudentParticipants computer science students) whether a hint about the difficulty of the code to be understood in a task description anchors participants in their own code comprehensibility ratings.
  Subjective code evaluations are a commonly used measure for how well a developer in a code comprehension study understood code. Accordingly, it is important to understand how robust these measures are to cognitive biases such as the anchoring effect.\\
  Our results show that participants are significantly influenced by the initial scenario description in their assessment of code comprehensibility. An initial hint of hard to understand code leads participants to assess the code as harder to understand than participants who received no hint or a hint of easy to understand code. This affects students and professionals alike.
  We discuss examples of design decisions and contextual factors in the conduct of code comprehension experiments that can induce an anchoring effect, and recommend the use of more robust comprehension measures in code comprehension studies to enhance the validity of results.
\end{abstract}

\ccsdesc[300]{Software and its engineering~Software organization and properties}
\ccsdesc[300]{Human-centered computing~Empirical studies in HCI}

\keywords{code comprehension, anchoring effect, empirical study design, software metrics}

\maketitle

\section{Introduction\label{sec:intro}}

Thinking about what you tell participants at the beginning of your study matters --- at least if your concern is to produce a study design with high validity and reduce the probability of biased results.
Consider the following scenario.
Caroline is a developer who wants to take part in an advertised study to research the influence of code comments on source code comprehensibility.
On site, the study leader explains the study process to her.
She would have to look at a source code snippet and rate its comprehensibility.
Caroline is a little nervous, but the study leader instinctively reassures her that the code will not be too difficult to understand.
The study leader has good intentions here, yet there might be unattended consequences for this action.

The reader might start seeing, at this point, the internal validity of the fictitious study threatened, since the assessment of code comprehensibility was most likely influenced at this moment by the words of the study leader.
The introduction to a study should strictly follow a predefined script. The Standards for Educational and Psychological Testing enlist several recommendations to assemble and present instructions to administer a test, including the instructions presented to test takers so that ``it is possible for others to replicate the administration conditions under which the data on reliability, validity [...] were obtained''~\cite{apa2014}[p. 90]. 
Additionally, by following a predefined script, experimenter expectancies can be avoided~\cite{Wohlin:2012:Experimentation}.
We share this sentiment, and yet, from the researchers' perspective, it does not always turn out to be that simple to control for any external influences on subjective assessments.

Subjective code comprehension assessments are a common measure in code comprehension studies because of their simplicity~\cite{Oliveira:2020:Evaluating}.
At the same time, we know from more than forty years of research on the anchoring effect~\cite{Tversky:1974:Anchoring,Furnham:2011:ReviewAnchoring} that even inconspicuous environmental factors are sufficient to influence people in their estimation~\cite{Critcher:2008:Incidental}.
The anchoring effect denotes that an initial value is insufficiently adjusted so that \enquote{different starting points yield different estimates, which are biased toward the initial values}~\cite{Tversky:1974:Anchoring}.
It is one of the most robust cognitive biases~\cite{Furnham:2011:ReviewAnchoring} and, in the context of software engineering, the most studied~\cite{Mohanani:2018:Cognitive}.

A recent study by~\citet{Wyrich:2021:Mind} showed that a single code comprehensibility metric next to the source code being evaluated is sufficient to significantly anchor developers in their evaluation of the code.
So far, however, we do not know whether these findings are also applicable to scenario descriptions presented to experimental participants \textit{before} the actual code comprehension task and whether prior information about the code snippet leads to anchoring participants in their code comprehension assessments.

For the design of code comprehension studies, confirmation of such an effect would imply that subjective ratings should only be used when contextual factors can be controlled for with a high degree of certainty for all participants, thus minimizing the risk of anchoring individual participants and biasing the experiment results.
In any case, additional insights on the influence of scenario descriptions on subjective code comprehension ratings provide a useful basis for design decisions, and can partially counteract existing uncertainty about what constitutes a good empirical study~\cite{Siegmund:2016:PastPresent,Siegmund:2015:Views}.

For these reasons, we conduct a controlled experiment with \totalNumbParticipants participants to investigate the following \textbf{research question:} \begin{displayquote}
Does specific information available in advance about a code snippet influence developers in their subjective assessment of the code's comprehensibility?
\end{displayquote}

\noindent The following section summarizes related work on the anchoring effect, in and outside of software engineering.
In section 3, we describe our research design, the underlying conceptual model, and how we analyzed the collected data.
Sections 4 and 5 present and discuss the results to the research question and its implications.
Section 6 concludes the paper.

\section{Related Work\label{sec:related}}

Cognitive biases refer to systematic deviations from optimal decisions and judgment, which can potentially jeopardize the success of a project~\cite{Ralph:2011:Debiasing}.
In a systematic mapping study,~\citet{Mohanani:2018:Cognitive} show that cognitive biases can be found in many forms in software engineering: 37 different cognitive biases were identified in 65 articles.  The most frequently studied in the SE literature is the anchoring effect.

In a recent field study,~\citet{Chattopadhyay:2020:Tale} found that development is often disrupted by cognitive biases, such that biased actions are significantly more likely to be reversed later.
Actions associated with cognitive biases of the \textit{fixation} category, which includes the anchoring effect, were most likely to be reversed.

When~\citet{Tversky:1974:Anchoring} investigated the anchoring effect, they demonstrated the effect by spinning a wheel of fortune with numbers between 0 and 100.
The resulting number influenced how high or low participants estimated the share of African countries in the United Nations.
Note that there is apparently no relationship between the number on a wheel of fortune and UN countries.

In the decades that followed, numerous other studies on the anchoring effect were carried out~\cite{Furnham:2011:ReviewAnchoring}, so that today we know, for example, that environmental anchors can be much more subtle without losing their effect~\cite{Critcher:2008:Incidental}. In addition, the effect is not limited to laymen, but experienced people are also anchored in corresponding contexts~\cite{Furnham:2011:ReviewAnchoring}.

It has also been shown that the anchor does not necessarily have to be a number.
For example,~\citet{Allen:2006:LittleHelp} found that SQL queries can serve as anchors, and while developers in a query formulation task are faster when modifying existing SQL queries instead of rewriting them from scratch, solutions were as well less accurate and overconfidence in the results increased.

While the anchoring effect generally receives ample attention, studies to better understand its actual influence and occurrence in the context of program comprehension experiments are still lacking.
Yet, this is precisely where these studies are needed.
Code comprehension and code comprehensibility are often measured in code comprehension studies by participants' subjective ratings~\cite{Oliveira:2020:Evaluating}.
Potential biases in these measures due to uncontrolled environmental factors pose a validity threat.

\citet{Wyrich:2021:Mind} showed in a recent experiment that displaying different values of a made-up code comprehensibility metric significantly anchored study participants in their subjective ratings of source code comprehensibility.
Participants were assigned to one of two groups and saw three code snippets one after the other, which they had to understand and evaluate in terms of their understandability.
All participants were shown a supposedly validated metric next to the code snippets, which however displayed a different value depending on the treatment group (either 4 or 8) and was intended to anchor the participants in this way.
The observed anchoring effect was significant ($p<.01$) with a large effect size ($d = -1.29$).

\begin{table*}[t]
  \caption{Comparison of experiment characteristics with the study by \citet{Wyrich:2021:Mind}}
  \label{tab:comparison}
  \begin{tabular}{lll}
    \toprule
    & \citet{Wyrich:2021:Mind} & Our study\\
    \midrule
    Scenario presentation & next to code snippet, metric only & prior to code snippet, metric and snippet details\\
    Metric values (scenarios) & 4, 8 & 3, 8, none\\
    Comprehension task & determine output and rate understandability & rate understandability\\
    Participants & 43 students & \numbStudentParticipants students, \numbProfessionalParticipants professionals\\
    Setting & onsite, code shown on screen & remote, code shown on screen\\
  \bottomrule
\end{tabular}
\end{table*}

The experiment by~\citet{Wyrich:2021:Mind} is closest to ours, but differs decidedly in some respects, which is why we do not refer to ours as a replication study.
\citet{Wyrich:2021:Mind} anchored the participants by displaying the anchor, i.e., the numeric value for the metric, during the code comprehension task, next to the code snippets, thus much more prominently than we do. We study the anchoring of experiment participants at the beginning of the study using a description of the expected code snippet, hence, displayed before showing the coding snippet.
Our choice is based on the limitation discussed by~\citet{Wyrich:2021:Mind} that showing the anchoring element during the code comprehension task might be unusual.
We sought to verify that the results still hold true when we were closer to realistic scenarios, some of which we discuss in this paper.

Other differences in design characteristics between the two studies are outlined in Table~\ref{tab:comparison}.
Nevertheless, similar to a \textit{reproduction} study\footnote{https://www.acm.org/publications/policies/artifact-review-badging}, we build on the work by~\citet{Wyrich:2021:Mind} to make both studies as comparable as possible and thus contribute to a common overall picture.

In summary, the anchoring effect is present in software engineering and its investigation follows not only current efforts to debias software engineering practice eventually~\cite{Ralph:2011:Debiasing,Mohanani:2018:Cognitive,Chattopadhyay:2020:Tale}, but also to identify confounding factors in the context of scientific studies and thus enable the design of valid studies.

\section{Methodology\label{sec:methodology}}

The goal of our study is to investigate the effect of the presence and content of information about a code snippet at the beginning of a scientific study on the participant's rating of the comprehensibility of that code snippet.
To this end, we formulated the research question given in the introduction.

\subsection{Research Design}

\begin{figure*}[t]
  \centering
  \includegraphics[width=\linewidth]{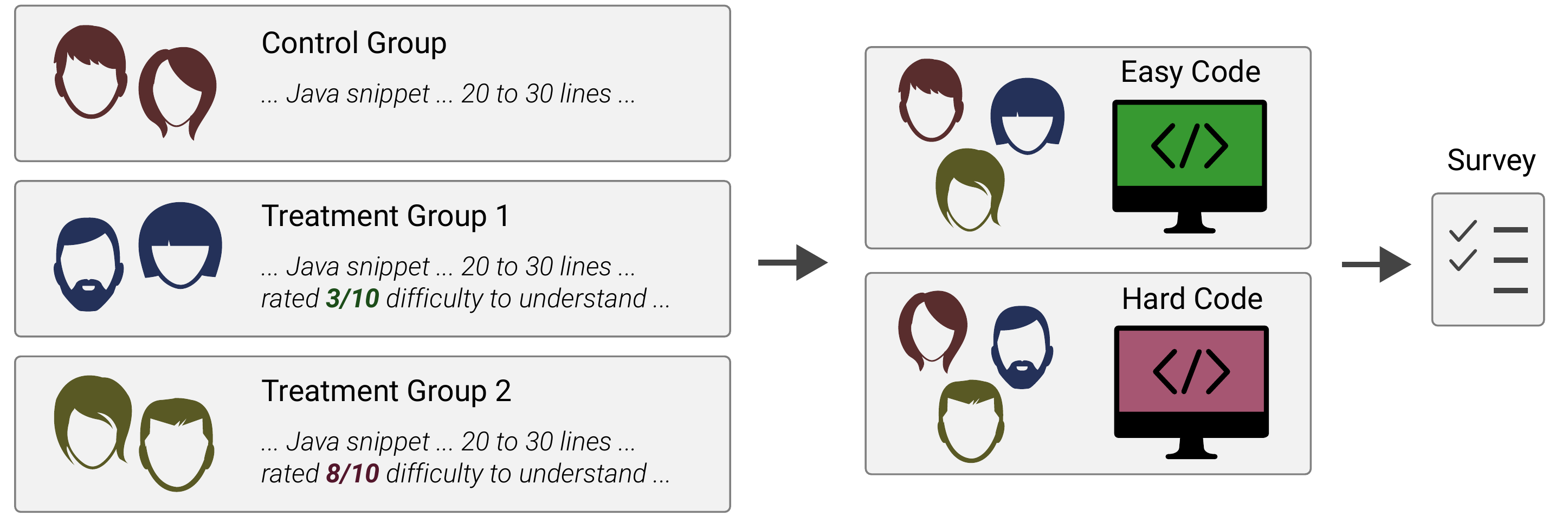}
  \caption{Schematic representation of the research design. Participants are randomly assigned to one of three groups that provide different information about the code snippet to be understood next. Each participant then randomly sees exactly one from a pool of two snippets, either an easy one or a difficult one. A survey that is the same for all concludes the experiment.}
  \Description{Schematic representation of the research design.}
  \label{fig:design}
\end{figure*}

We conducted a controlled and randomized between-subjects experiment with 3x2 factorial design.
Each participant was assigned to one of three scenario groups and had to understand one of two code snippets.
A schematic representation of the research design is provided in Fig.~\ref{fig:design}.

The experiment took place online via a self-hosted instance of LimeSurvey, and participation was entirely anonymous.
Participants first confirmed that their consent to participate was informed and agreed to the data and privacy policy.
On the next page, participants saw a description of the task that awaited them in the next step.
Depending on the randomly assigned group, this description included information about the code comprehensibility rating of a fictitious expert system.
Then, each participant had to understand one randomly selected code snippet and provide their code comprehensibility rating.
Finally, a demographic data questionnaire concluded the study.

\subsection{Experimental Materials}

\subsubsection{Scenarios}
All participants saw a textual scenario description of the expected code comprehension task.
The wording of this description was the same for all three groups, but one of three paragraphs was not shown to the control group, and there was a hint to Treatment Groups 1 and 2 for a value of either 3 (easy) or an 8 (hard) as a comprehensibility score by the expert system.
The description was as follows:

\begin{displayquote}
You will look at a Java code snippet in a bit. The only information that we
provide you about the snippet is the following: The code snippet is between
20 and 30 lines long and tests a String, a sequence of characters, for a
criterion.\\

$\langle\langle$Begin Treatment Group text snippet$\rangle\rangle$\\
We developed an expert system to rate the understandability of code based
on multiple metrics. This system rated the code snippet you will look at in a
few moments as 

$\langle\langle$Treatment Group 1$\rangle\rangle$ \textbf{a 3 out of 10.}

$\langle\langle$Treatment Group 2$\rangle\rangle$ \textbf{an 8 out of 10.}

The system uses a scale from 1 to 10,
where 1 is very easy and 10 is very hard to understand.\\
$\langle\langle$ End Treatment Group text snippet$\rangle\rangle$\\

Your only task is to judge its understandability on a scale from 1 to 10, where 1 is very easy to understand and 10 is very hard to understand. The code is fully functioning and bug-free. You will have unlimited time to look at the code snippet. Feel free to rate the
code snippet whenever you think you have an adequate impression to judge
its understandability.
\end{displayquote}

\subsubsection{Code Snippets}
Since the influence of a scenario description on anchoring in source code comprehensibility ratings might depend on the actual difficulty of the code snippet, we were interested in studying both an easy and a hard snippet in the study.

Every developer has a slightly different idea of how understandable certain code is. Therefore, we pre-selected five Java code snippets and invited eight software developers to assess their comprehensibility.
The pre-selected snippets all met the criteria that no domain knowledge is necessary for understanding and that they are neither too long to be displayed in full on a screen nor too trivial to be understood after just a few seconds.
In pre-selecting functions that are rather complex, we, like~\citet{Wyrich:2021:Mind}, used the cognitive complexity metric~\cite{Campbell:2018:CogComplexity}, which has been shown to correlate in particular with subjective evaluations by developers~\cite{Baron:2020:CogComplexity}.

Through this preliminary evaluation, we were able to identify one easy and one difficult code snippet that we subsequently used in our study.
The easy code snippet is a method from the apache commons-lang \verb|StringUtils| class and checks if a given character sequence contains both uppercase and lowercase characters.
The difficult code snippet is the solution to a coding challenge~\cite{Wyrich:2019:Theory} to find the longest palindromic substring within a string.
Both code snippets are included in the supplemental materials.

\subsubsection{Questionnaire}
The experiment was completed by a short demographic questionnaire, which asked about the current main occupation of the participant.
The choices were \textit{Student}, \textit{IT professional}, \textit{Researcher} and \textit{Other} (with the possibility to specify the occupation).
We then clarified the specific intent of our study and again listed ways to contact us with potential questions or comments.

\subsection{Participants\label{ssec:participants}}
We invited a convenience sample of software professionals and computer science university students to take part in our study.
To disseminate the invitation, we used social media and asked personal contacts to draw attention to the study in their software companies.
Computer science university students (software engineering curriculum, BSc and MSc) were asked to participate in the study, for example to fulfill course requirements to participate in scientific studies.

We assured the participants of anonymity, and they could drop out of the study at any time or not participate at all and still fulfill their requirements.
The only requirement for participation was a basic understanding of the Java programming language.

We encouraged participation with the low amount of time commitment of 10 to 15 minutes for the whole study.
Furthermore, we pledged to donate €5 to a good cause for every participant among the software professionals that completed the study.
Subjects could choose between three charity projects on different topics, or split the donation evenly among the three projects.

Following the goal-setting theory of motivation~\cite{Locke:2002:GoalSetting}, we also invited participants with the clear goal of assessing the comprehensibility of a particular code snippet.
Setting such a specific and challenging goal may lead to increased effort and persistence, which is desirable for completing the study.
Our purpose in doing so was to pique the interest of developers who want to demonstrate their ability to understand code.

\subsection{Conceptual Model\label{sec:concept}}

\begin{figure}[t]
  \centering
  \includegraphics[width=\linewidth]{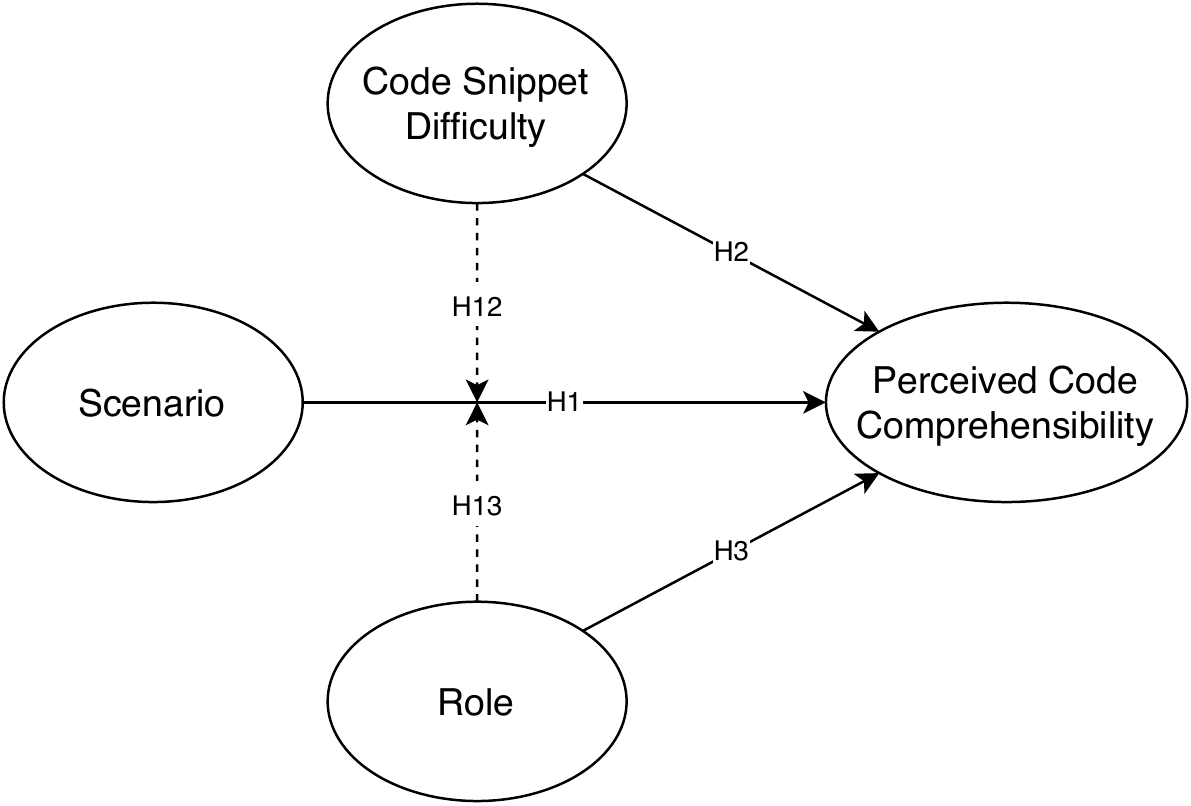}
  \caption{Conceptual model for the study. Dashed arrows represent moderators.}
  \label{fig:model}
\end{figure}

We aim to build on the study by~\citet{Wyrich:2021:Mind} by developing the conceptual model of Figure~\ref{fig:model}, which summarizes all variables and their hypothesized relationships graphically.
We investigate, like \citet{Wyrich:2021:Mind}, whether the anchoring effect is confirmed in the form that a \textit{Scenario} description, which controls for information presented at the beginning of the study about the code snippet to be understood, influences \textit{Perceived Code Comprehensibility} \textbf{(\texttt{H1})}.
In our model, we introduce two other factors that we theorize to have an effect on Perceived Code Comprehensibility. The \textit{Code Snippet Difficulty} will influence the Perceived Code Comprehensibility \textbf{(\texttt{H2})}.
\citet{Wyrich:2021:Mind} selected the code snippets for their study to be of comparable complexity to control for the potential influence of snippet complexity in this way.
We appreciate this design decision, but we want to understand the extent to which an easy or a hard task affects the Perceived Code Comprehensibility, and we argue that this influence will be stronger than the one provided by the Scenario.

\begin{table}[t]
\caption{Assignable values for the variables of the model.}
\label{tab:variables}
\begin{tabularx}{\columnwidth}{l l}
\toprule
Variable & Values \\ \midrule
Scenario & \{baseline, easy, hard\} \\
Code Snippet Difficulty & \{easy, hard\} \\
Role & \{student, professional\} \\
Perceived Code Comprehensibility & [1.. 10]\\ 
\bottomrule
\end{tabularx}
\end{table}

Furthermore,~\citet{Wyrich:2021:Mind} had only students as participants in their study.
Much discussion has happened in the literature on differences between students and professionals~\cite{Feldt:2018:FourCommentaries} in terms of productivity, performance, and software quality, with some claiming or finding that there are little to none~\cite[e.g.]{Salman:2015:students}, others that there are~\cite[e.g.]{Sjoberg:2002:Realistic}.
\citet{Wyrich:2021:Mind} themselves discuss this circumstance as a potential limitation of their study, although literature suggest that the anchoring effect is not restricted to inexperienced people~\cite{Tversky:1974:Anchoring, Furnham:2011:ReviewAnchoring}.
We avoid any debate and investigate whether a \textit{Role} influences the Perceived Code Comprehensibility \textbf{(\texttt{H3})}.

Finally, given the absence of prior literature, we want to test for the influence that these factors have with each other.
The interaction between Scenario and Code Snippet Difficulty \textbf{(\texttt{H12})} as well as the interaction between Scenario and Role \textbf{(\texttt{H13})} are further modeled as \textit{moderators}\footnote{A variable \textit{w} is called a moderator when the relationship between other two variables \textit{A} and \textit{B} is influenced by \textit{w}, and it is modelled as interaction effect~\cite{cohen2013}.} on the influence that Scenario has on Perceived Code Comprehensibility.
Investigating the moderators will enable us to better characterize a potential anchoring effect of Scenario on Perceived Code Comprehensibility.

Table~\ref{tab:variables} provides a summary of the values that can be assigned to each of the four variables in the conceptual model.
Due to the small number of six self-identified researchers, we decided to combine them with the 44 software professionals into one group for the analysis. We still consider that a meaningful distinction can be made between students and professionals.
Nevertheless, we will discuss the potential consequences of this design decision in~\ref{ssec:limitations}.

\subsection{Analysis Procedure}\label{ssec:analysis-procedure}

The data violated at least one assumption of most of the commonly used modelling techniques\footnote{Including linear regression models, methods from the various ANOVA families with applied data transformation techniques, and ordinal logistic regression methods. The latter could not be applied because of proportional odds violation.}, and it also presented evidence for non-normality on the dependent variable (Shapiro-Wilk Test, $W = 0.97$, $p < .00001$).

We thus opt for a Partial Least Squares Structural Equation Model (PLS-SEM). PLS-SEM was recently introduced to the discipline by~\citet{Russo:2021:PLS} as part of the SEM statistical technique family for causal-predictive approaches in the behavioral sciences. 
We direct readers to~\citeauthor{Russo:2021:PLS}'s work~\cite{Russo:2021:PLS} for an introduction and overview but, in short, PLS-SEM is suited for testing a theoretical framework from a prediction perspective, when the structural model is complex, and when distribution issues are a concern~\cite{hair2019}.

We model the factors of Figure~\ref{fig:model}, as defined in~\ref{sec:concept}, as a PLS-SEM in \textit{R 4.1.2}~\cite{R2021} and \textit{SEMinR 2.2.1}~\cite{Seminr2021}. The output will provide us with statistics on the predictive power and significance of the relationship between the factors\footnote{Readers familiar with SEM will notice that we are applying PLS-SEM as an analysis technique to estimate single-item indicators. That is, we rely on PLS-SEM robustness to perform a ``regression job'' instead of using the technique for its intended psychometric purposes. We elaborate on these issues in the limitations section.}.

\section{Results\label{sec:results}}

We recruited \totalNumbParticipants participants, of which \numbStudentParticipants were students and \numbProfessionalParticipants were professional software developers (see Section~\ref{ssec:participants}).

Table~\ref{tab:descriptive} provides descriptive statistics for the Perceived Code Comprehensibility (PCC) grouped by factorial assignment condition and role.
Students and professionals were overall very close or identical in their median PCC for the same scenario and code snippet.
The code snippet we considered easy was actually perceived by participants as easier to understand than the snippet predicted to be perceived harder to understand, regardless of scenario and role.
Participants provided the highest median PCC ratings for the combination of hard scenario and hard code snippet (a median value of 7 and 7.5 from students and professionals).
For the easy code snippet, the scenario seems to have had a smaller overall impact on the PCC ratings.

\begin{table}[hbt]
\caption{Descriptive statistics and group assignment for the study. CSD = Code Snippet Difficulty, n = group size, PCC = Perceived Code Comprehensibility, M = mean, SD = standard deviation, Mdn = median.}
\label{tab:descriptive}
\begin{tabularx}{\columnwidth}{p{0.15cm} l l r r r r}
\toprule
 & CSD  & Role         & n   & PCC M & PCC SD & PCC Mdn \\ \midrule
\multicolumn{7}{l}{\textbf{\textit{Scenario: baseline}}}\vspace{0.1cm} \\
         & easy &              &     &       &        &         \\
         &      & student      & 35  & 4.0   & 3.33   & 2       \\ 
         &      & professional & 10  & 3.3   & 2.79   & 2       \\
         & hard                                                              &              &   &       &        &         \\
         &                                                                   & student      & 32                     & 5.25                       & 2.05                        & 5                            \\
         &                                                                   & professional & 10                     & 5.6                        & 2.88                        & 6                            \\ \\
\multicolumn{7}{l}{\textbf{\textit{Scenario: easy}}}\vspace{0.1cm} \\
         & easy                                                              &              &   &       &        &         \\
         &                                                                   & student      & 25                     & 2.56                       & 1.94                        & 2                            \\
         &                                                                   & professional & 9                      & 2.89                       & 2.67                        & 2                          \\
         & hard                                                              &              &   &       &        &         \\
         &                                                                   & student      & 33                     & 4.67                       & 1.49                        & 4                            \\ 
         &                                                                   & professional & 5                      & 6.0                        & 2.65                        & 7                            \\ \\
\multicolumn{7}{l}{\textbf{\textit{Scenario: hard}}}\vspace{0.1cm} \\
         & easy                                                              &              &   &       &        &         \\ 
         &                                                                   & student      & 32                     & 3.25                       & 2.90                        & 2                            \\
         &                                                                   & professional & 2                      & 1.0                        & 0.0                         & 1                            \\
         & hard                                                              &              &   &       &        &         \\ 
         &                                                                   & student      & 49                     & 6.22                       & 2.05                        & 7                            \\ 
         &                                                                   & professional & 14                     & 6.86                       & 2.35                        & 7.5                            \\ \bottomrule
\end{tabularx}
\end{table}

In Table~\ref{tab:coefficients} we show the results of the statistical analysis for each hypothesis.
We find a significant path from Scenario to Perceived Code Comprehensibility (path coefficient\footnote{Path coefficients are expressed as standardized regression coefficients in terms of standard deviations; see, e.g.,~\cite{hair2011}.} $\beta = 0.165, p<.001$) confirming the presence of the anchoring effect.
Code snippet difficulty had the expected strong influence on PCC ($\beta = 0.418, p<.001$), but, furthermore, seems not to be a moderator of the relationship between Scenario and PCC ($\beta = 0.076, p>.10$).
The paths from Role to PCC and the moderating effect of Role on the path from Scenario to PCC were insignificant.
Thus, our results support H1 and H2; they do not support H12, H3 and H13.

\begin{table*}[hbt]
\caption{Estimated and bootstrapped path coefficients with 95\% CI, model explanatory power expressed as R\textasciicircum{}2 and adjusted R\textasciicircum{}2.\newline$* = p \leq .10$, $** = p \leq .01$, $*** = p \leq .001$.}
\label{tab:coefficients}
\begin{tabularx}{\textwidth}{l l l l l r}
\toprule
                                       & Perceived Code Comprehensibility & Bootstrap Mean      & Bootstrap SD & T Stat.            & 95\% CI             \\
\midrule
H1: Scenario                           & \phantom{-}0.165***              & \phantom{-}0.169    & 0.049        & \phantom{-}3.357   & {[}0.071, 0.265{]}  \\ 
H2: Code Snippet Difficulty            & \phantom{-}0.418***              & \phantom{-}0.418    & 0.063        & \phantom{-}6.673   & {[}0.293, 0.541{]}  \\ 
H3: Role                               & \phantom{-}0.033                 & \phantom{-}0.031    & 0.058        & \phantom{-}0.564   & {[}-0.08, 0.148{]}  \\ 
H12: Scenario*Code Snippet Difficulty  & \phantom{-}0.076                 & \phantom{-}0.074    & 0.055        & \phantom{-}1.384   & {[}-0.034, 0.181{]} \\ 
H13: Scenario*Role                     & -0.013                           & -0.017              & 0.058        & -0.228             & {[}-0.124, 0.096{]} \\ 
R\textasciicircum{}2                   & \phantom{-}0.223                 &                     &              &                    &          \\ 
AdjR\textasciicircum{}2                & \phantom{-}0.208                 &                     &              &                    &          \\
\bottomrule
\end{tabularx}
\end{table*}

The model explains $R^2=.223$ of the variance in Perceived Code Comprehensibility.
Bootstrapped heterotrait-monotrait ratio of correlations (HTMT) are all below zero, bootstrapped loadings and weights are all approximately 1.00, all variance inflation factors (VIF) are 1.00, and all reliability coefficients are 1.00. All this is expected with single item constructs that are assumed to be fully independent.

\section{Discussion\label{sec:discussion}}

We can answer the research question whether specific information available in advance about a code snippet influences developers in their subjective assessment of code's comprehensibility as follows: The anchoring effect is, once again, significant.
Information about the expected complexity of a code snippet to be understood influences developers in their supposedly independent code evaluations.
Students and professionals are equally affected, which is in line with the investigation of~\citet{Wyrich:2021:Mind} who measured programming experiences instead of role, and found programming experience to likely not play a role in anchoring.
The finding is further in line with the body of literature on the anchoring effect, which states that the anchoring effect is not limited to laymen or those inexperienced in an activity~\cite{Tversky:1974:Anchoring,Furnham:2011:ReviewAnchoring}.

The choice of code snippet also has a significant impact on perceived code comprehensibility.
While this is not too surprising, nor is it the central point of our work, it is still good to have data points that show that the choice of code snippets for code comprehension studies should not be underestimated.
Studies seeking to ensure that code snippets of different tasks are of comparable difficulty should invest effort in, e.g., a pilot study to evaluate appropriate snippets.

What is interesting for the characterization of the observed anchoring effect, however, is that the complexity of a code snippet did not have an influence on the strength of the anchoring effect (H12).
Based on our results, we suspect that a snippet must be complex enough so that PCC ratings do not concentrate too much on the lower end of the scale.
Apart from that, the actual complexity of a snippet does not seem to be too important for anchoring.

\subsection{Implications}
\label{ssec:implications}

We know, regarding the anchoring effect, that the specific anchor can take many forms and does not necessarily always have anything to do with the situation being assessed~\cite{Tversky:1974:Anchoring,Furnham:2011:ReviewAnchoring}. 
When designing and conducting code comprehension studies, researchers have a lot of freedom and just as much potential to inadvertently set anchors.
In the introduction, we described a scenario in which the instructor mentioned, to encourage the participant, that the code snippets to be assessed were not too difficult to understand.
Similarly, however, even in an online experiment without a human instructor, many examples of potential anchors can be found.

For example, explicitly mentioning in a study description that it is a study for novice programmers could lead to code snippets being rated as easier to understand.
Communicated time limits for the code comprehension tasks can convey to participants how complex the tasks are supposed to be.
A validation study of a code comprehensibility metric that displays the metric to developers and asks whether they agree with it or how they would assess the code instead already anchored them in their own judgments, as our study and that of~\citet{Wyrich:2021:Mind} have empirically demonstrated.

One can easily find further examples in which study participants would be anchored.
We see great research potential to empirically investigate these scenarios and to find solutions for affected studies, which for example cannot disregard subjective evaluations because they are relevant for their research intentions.
Yet, for all study designs that allow to dispense with subjective assessments, we recommend more reliable and objective measurements to draw conclusions about how well a developer has understood code.
Corresponding tasks and measures are available for this purpose~\cite{Oliveira:2020:Evaluating,Fakhoury:2018:Objective,Feitelson:2021:Considerations}.

\subsection{Limitations}
\label{ssec:limitations}

With our design, we were able to overcome a number of limitations discussed in the work by~\citet{Wyrich:2021:Mind}.
Our experiment had a much larger and more heterogeneous sample of developers, a less prominent presentation of the anchor, and we had a control group that allowed us to measure the perceived code comprehensibility for participants who were not explicitly anchored.
Still, the results of our study should be seen in the light of some limitations.

A confounding factor for the assessment of the code snippets' understandability might be diverse understandings of what constitutes comprehensible code between the participants.
We discussed, while designing the study, whether to provide a definition of understandability at the beginning of the survey.
However, since there exists neither an agreement in the literature nor does it seem realistic that developers all share the same view on understandability, we decided against it.
It might be possible that participants in one group share similar views on what constitutes understandable code, while participants in other groups differ. On the other hand, the randomized assignment of participants to scenario and code snippet should have mitigated this threat.

Regarding the generalizability of our results, we would like to emphasize that the scope clearly lies on the evaluation of individual code snippets and participants applied a comprehension process commonly referred to as bottom-up code comprehension~\cite{OBrien:2004:BottomUp}.
We see much value in reproductions of our experiment with larger software systems to be assessed.

We conducted our experiment remotely to make the study accessible to as many developers as possible and to minimize contacts due to the pandemic.
The context in which the participants took part in the study could therefore not be controlled, which could have caused individual participants to be distracted during the conduct of the study or not to complete the study conscientiously.
We do not know what influence a potential distraction can have on a code evaluation, but at least we think that the activity can be resumed after an interruption.
Participants who spent less than 20 seconds viewing and rating the code snippet's understandability were excluded from the analysis (average time for this task was around four minutes; a total of 10 participants were excluded based on this criterion).

Apart from this, running the experiment remotely worked well, and we are pleased to have recruited 50 software professionals as participants in addition to students.
We have previously disclosed that this experiment group included both software professionals and six participants who identified themselves as researchers. This could be seen as a threat to construct validity, since software development experience might be less pronounced among software engineering researchers than among full-time software engineering professionals.

We assume that researchers in computer science, who hold either a related MSc or a PhD, have comparable experience compared to software professionals to be able to deal with our tasks.
Yet, we repeated the statistical analysis without the six researchers, and there was no change in the significance of the results. The path coefficients for the significant hypotheses would be $\beta\_H_1=0.161$ and $\beta\_H_2=0.408$ (change $\leq$ 0.10).

Further, when designing the experiment, we took care to minimize the time required for participation to achieve a correspondingly high response rate.
This is the practical reason why we did not measure actual code comprehension with additional tasks. Furthermore,~\citet{Wyrich:2021:Mind}~found evidence for a strong anchoring effect in perceived code comprehensibility ratings, but also that actual code comprehension was not affected by the anchoring effect. We assume that by our random assignment of a sample six times larger than that of~\citet{Wyrich:2021:Mind} the absence of effect anchoring/actual understanding still holds.

As anticipated in Section~\ref{ssec:analysis-procedure}, we make use of PLS-SEM in ways that reduce it to a simpler regression modelling tool with single-item constructs. That is, we use PLS-SEM to understand the relationship between variables rather than typical reflective and/or formative constructs in structural and measurement models. As disclosed in the same section, we were driven to this choice because we preferred not to rely on robustness against assumption violations of other analysis and regression tools. PLS-SEM has just a few assumptions that our data did not violate, and it suited our needs. 

Furthermore, we tested our results with an alternative model specification and report coherent results. We fit an equivalent multilevel mixed effects model with \textit{lme4 1.1-27.1 }~\cite{Bates2015}. Relying on its robustness against non-randomness of residuals, which is the case of our data, we find that 
\begin{enumerate*}
\item the same path coefficients that are found as significant in our PLS-SEM are also significant in the multilevel mixed effects model,
\item the model provides comparable explanatory power ($R^2 = .22$ and adjusted $R^2 = .22$)
\end{enumerate*}.
We are thus confident in our chosen analysis tools and its results.
For elaborating on the limitations, we opted for multilevel mixed effects models based on~\citeauthor{Clark2016}'s report~\cite{Clark2016} on their positive comparability with SEM. We still opted not to use multilevel mixed effects models for our main analysis because of cautiousness: the residuals are not randomly distributed.

The other related issue lies in using single item scales to represent our underlying constructs. PLS-SEM allows for single item constructs,  and it is also used this way in information systems research~\cite{petrescu2013}. This use is, however, encouraged only when no other alternatives are available since psychometric properties of single item constructs are subject of debate of providing low psychometric validity and low reliability~\cite{ringle2012,petrescu2013}. 
We argue that we fall within the recommendation of employing PLS-SEM for single item constructs. First, our independent variables are experimental groups represented by either two or three ordinal or categorical options. Furthermore, they provide highly observable validity and reliability (e.g., \texttt{role = student} indeed represents students, and \texttt{scenario = easy} really represents the easy scenario). 
Only the dependent variable, Perceived Code Comprehensibility, is operationalized by a single item, and the results might indeed suffer from low psychometric validity and reliability.

The Perceived Code Comprehensibility is a variable of which scale we inherited by reproducing~\citeauthor{Wyrich:2021:Mind}'s study~\cite{Wyrich:2021:Mind}. This was a deliberate choice to be able to compare results and build on their study. A psychometrically validated multi-item scale to assess the Perceived Code Comprehensibility would allow for a richer and better explanation for the variance therein and offer (more) valid and reliable interpretation of its values. Such a scale, alas, does not exist, to the best of our knowledge. We report on this issue that is shared by most, if not all, studies that investigate Perceived Code Comprehensibility and call for future research on more robust ways to assess it.

\section{Conclusion\label{sec:conclusion}}

The conduction of program comprehension studies with human participants all usually start in the same way: with an introduction of the participants to the study context.
We have shown in a controlled experiment with \totalNumbParticipants participants that a subtle cue about the difficulty of the code to be understood anchors participants in their subjective assessments of code comprehensibility.
This result is consistent with prior research and is one reason to design code comprehension studies somewhat differently in the future, as many studies today still rely on subjective assessments of source code as a proxy for how well a participant understood code.
Such a comprehension measure is tempting because it provides an indicator of code understanding in a relatively time efficient manner.
However, it is just as easy for this assessment to be biased by contextual factors, which is why we recommend the use of more robust measures such as the correctness of comprehension questions.

Future work should explore ways to predict and mitigate the anchoring effect as a confounding factor in code comprehension studies interested in perceived code comprehensibility.
Debiasing software engineering through empirical studies is only possible if we also debias empirical studies themselves.
On a related note, we think it is worthwhile to investigate other, even more subtle, potential anchors.
In section~\ref{ssec:implications}, we have discussed a few examples that could affect not only code comprehension studies, but empirical studies in software engineering in general.

\section{Data Availability\label{sec:data}}

We publicly provide the code snippets used, the data collected, and the analysis script as supplementary material at \url{https://doi.org/10.5281/zenodo.5877313}.

\begin{acks}
We are grateful for the many participants who volunteered to take part in our study. We thank three anonymous reviewers for their constructive and insightful feedback.
\end{acks}

\bibliographystyle{ACM-Reference-Format}
\bibliography{references}

\end{document}